\documentclass[reprint, amsmath, amssymb, aps, prb, floatfix]{revtex4-1}

\usepackage{graphicx}
\usepackage{dcolumn}
\usepackage{bm}
\usepackage[normalem]{ulem}
\usepackage{hyperref}
\hypersetup{colorlinks = true, citecolor = blue, breaklinks = true}
\usepackage{balance}

\usepackage{mhchem}
\usepackage[resetlabels]{multibib}
\newcites{SI}{References}

\begin{document}

\title{The role of metastability in enhancing water-oxidation activity}
\date{\today}
\author{Nathalie Vonr\"uti}
\affiliation{Department of Chemistry and Biochemistry, University of Bern, Freiestrasse 3, CH-3012 Bern, Switzerland}
\author{Ulrich Aschauer}
\affiliation{Department of Chemistry and Biochemistry, University of Bern, Freiestrasse 3, CH-3012 Bern, Switzerland}

\begin{abstract}
While metastability enhanced water-oxidation activity was experimentally reported, the reason behind this effect is still unclear. We determine here, using density functional theory calculations, oxygen evolution reaction overpotentials for a variety of defective (001) surfaces of three different perovskite materials. For all three, we find a large range of overpotentials for different reaction sites including also overpotentials at the top of the activity volcano. Assuming that these sites dominate the apparent catalytic activity, this implies that a large number of geometrically different reaction sites, as they occur when a catalyst is operated at the border of its stability conditions, can lead to a strong enhancement of the apparent activity. This also implies that a pre-treatment of the catalyst creating a variety of different reactive sites could lead to superior catalytic activities for thermodynamically stable materials.
\end{abstract}

\maketitle

\section*{Introduction}

Hydrogen fuel production by (photo)electrochemical water splitting has been intensively studied as a route to convert electrical or solar energy to chemical energy. The bottleneck in (photo)electrochemical water splitting is the oxygen evolution reaction (OER). Previous studies of metal\cite{rossmeisl2005electrolysis} and oxide\cite{man2011universality} catalysts proposed a likely OER reaction mechanism to consist of four consecutive proton-coupled one-electron transfer (PCET) steps with reaction intermediates *\ce{OH}, *\ce{O} and *\ce{OOH} as shown in equations \ref{eq:1}-\ref{eq:4}.
\allowdisplaybreaks
\begin{align}
	&*         &+ \ce{H2O(l)} &\rightarrow *\ce{OH}  &             &+ \ce{H+} + \ce{e-}\label{eq:1}\\
	&*\ce{OH}  &              &\rightarrow *\ce{O}   &             &+ \ce{H+} + \ce{e-}\label{eq:2}\\
	&*\ce{O}   &+ \ce{H2O(l)} &\rightarrow *\ce{OOH} &             &+ \ce{H+} + \ce{e-}\label{eq:3}\\
	&*\ce{OOH} &              &\rightarrow *         &+ \ce{O2(g)} &+ \ce{H+} + \ce{e-}\label{eq:4}
\end{align}

Several experimental studies found an inverse correlation between the activity and the stability of catalyst materials for the OER. Markovic and co-workers investigated activity-stability trends of the OER for crystalline and amorphous \ce{RuO2}\cite{chang2015activity, danilovic2014activity} as well as for different surface orientations of the perovskite \ce{SrRuO3}\cite{chang2014functional}. Based on the concentration of dissolved metal ions, they found a higher activity for less stable surfaces of both materials and concluded that a highly active OER catalyst should balance activity and stability. A mechanistic understanding of a catalyst's stability in enhancing the OER activity is however still elusive and actively debated. It was suggested that the OER activity is controlled by the density of surface defects rather than by the binding energy of the reaction intermediates on the perfect surface and that it may be linked to changes in oxidation state of \ce{Ru} ions close to defects\cite{chang2014functional}. A very recent study on \ce{RuO2} thin films did, however, not find such a fundamental relationship between a material's stability and its activity\cite{roy2018trends}. While a correlation between the dissolution and the activity for different surface orientations was initially observed, it was also shown that the active OER sites are decoupled from the fastest-corroding \ce{Ru} sites\cite{roy2018trends}.

To fully understand the activity-stability relationship one would ideally study the OER on specific reaction sites of defective dissolving materials. Experimentally it is, however, very challenging to characterize the geometry and activity of specific reaction sites since spectroscopic methods probe the average state of a catalyst. Computationally, one could readily calculate the overpotential and hence the activity of a given reaction site with a specific local geometry, but there exists no well-established method to determine the surface structure of a material with a low stability under OER conditions. For materials with a thermodynamically stable bulk, reasonable surface structures can be determined from the surface energy as a function of the applied potential and the pH by means of so-called surface Pourbaix diagrams (SPD)\cite{rong2015abinitio}. For thermodynamically unstable materials, on the other hand, one can only investigate the stability of the bulk with respect to competing phases\cite{kim2017unraveling, singh2017electrochemical}, while a SPD would yield the surface with the minimal number of atoms as most stable, indicating that the thermodynamic equilibrium is the fully dissolved state. Nevertheless, in practice the dissolution may be kinetically slow and still allow these materials to be used as catalysts\cite{kim2017unraveling}. We refer here to such materials, that are thermodynamically unstable under OER conditions and for which we therefore cannot determine their surface structure by means of SPDs, as metastable materials.

The aim of the present work is to obtain a density functional theory (DFT) description of the OER on surfaces of metastable materials and to reveal how metastability can increase the performance of a catalyst. Computational studies on different oxides have, so far, mainly compared the OER activity of ideal stoichiometric, clean\cite{montoya2018trends} or adsorbate-covered\cite{montoya2014theoretical}, low-index surfaces which are unlikely to be good approximations for surfaces of a metastable material that may slowly dissolve under OER conditions\cite{kim2017unraveling}. Few studies reported the theoretical activity for reaction sites in the vicinity of point defects, steps or kinks\cite{rong2015abinitio, dickens2017theoretical, dickens2019electronic, seitz2016highly}, however without linking such structures to metastability. Our approach to account for metastability is to study the OER on a large variety of different reaction sites by sampling many defective surface structures that should be representative for a slowly dissolving metastable material. We study the effect of metastability on the OER for three chemically very different perovskite structured catalysts, which enables us to distinguish between effects resulting from intrinsic material properties and effects of specific reaction sites. We compare \ce{LaTiO2N} (LTON) and \ce{SrRuO3} (SRO), which differ by their $d^0$ respectively $d^4$ $d$-electron occupation and their insulating respectively metallic character. LTON is the best studied member of the family of perovskite oxynitrides that by virtue of their smaller band gap than oxides are considered promising photocatalysts\cite{kasahara2003latio3}. Calculations\cite{castelli2014calculated} and experimentally observed nitrogen loss\cite{kasahara2002photoreactions} indicate that LTON is metastable under OER conditions and SRO was also reported to be metastable\cite{chang2014functional, kim2017unraveling}. As these two materials differ not only in their metallic respectively insulating behaviour but also in their cation composition, we complement the set of materials with \ce{SrTiO3} (STO), which shares properties with both LTON and SRO: like SRO it is a pure oxide but like LTON it is a $d^0$ insulator. Moreover, it shares the same A and B site with SRO and LTON respectively but unlike the other two materials STO was not reported to be metastable.

\section*{Computational approach}

\begin{figure*}
 \centering
 \includegraphics{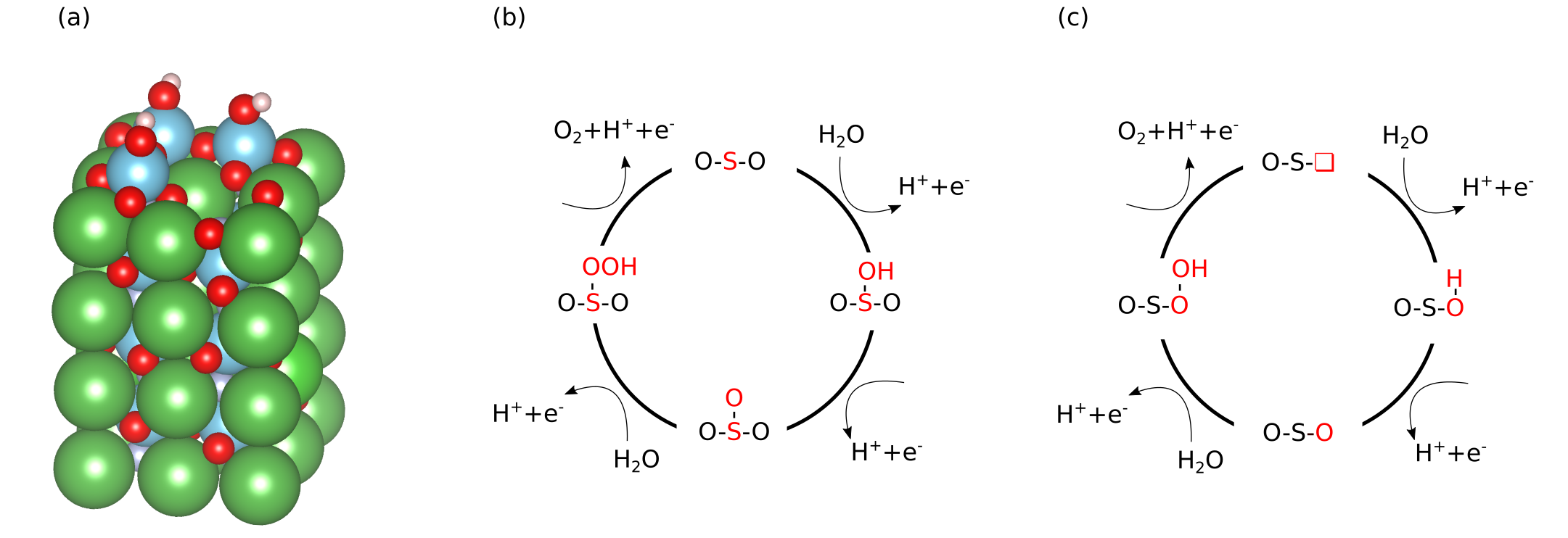}
 \caption{Surface structure example and catalytic cycles: (a) Example of a \textit{trans} \ce{LaTiO2N} slab with one Ti vacancy and the Ti atoms in the top layer covered with OH. Colour code: La = green, Ti = blue, O = red, H = white. (b) Conventional OER reaction mechanisms and (c) OER mechanism involving lattice oxygen.}
 \label{fig:fig1}
\end{figure*}

For all three materials we investigate both the \ce{AO} and \ce{BO2} terminated (001) surface, assuming that for a metastable or unstable material both terminations will be exposed as dissolution proceeds. For the oxynitride we also have to consider the anion order, which is different in the surface and the bulk: \ce{LaTiO2N} assumes a \textit{cis} anion order in the bulk to maximize the overlap between the N 2\textit{p} and the Ti 3\textit{d} orbitals\cite{yang2010anion} but the (001) surface prefers a non-polar \textit{trans} anion order\cite{ninova2017surface}. We therefore perform calculations on the \textit{trans} surface but to account for the metastability-induced dissolution of the thin \textit{trans} layer\cite{ninova2019anion} also investigate the OER on a \textit{cis} ordered surface. As we find no significant difference between the two anion orders regarding OER reaction free energies, we do however not distinguish between them in our results.

Our stoichiometric and clean surface slabs contain four surface cation sites and we create defective surfaces with between one and four nitrogen, titanium, ruthenium, lanthanum or strontium vacancies in the surface layer (see Fig. \ref{fig:fig1}a for an example). We do not explicitly consider oxygen vacancies, since these are unlikely to exist on the surface under oxidizing OER conditions\cite{montoya2014theoretical,ouhbi2019nitrogen}. Occasionally, during surface relaxation - in particular for high cation-vacancy concentrations - either \ce{O2} or \ce{N2} recombine and desorb. In these cases, we stop the calculation and remove the desorbed molecules before continuing the calculation, which can in effect lead to oxygen vacancies. Since potentials relevant for (photo)electrochemical water splitting often lead to formation of oxidising surface species\cite{montoya2014theoretical}, we also cover surface metal sites with either zero or one monolayer of \ce{OH} or \ce{O} in top position in the case of \ce{Ti} and \ce{Ru} as well as in the bridge position for \ce{La} and \ce{Sr} (see Fig. \ref{fig:fig1}a, as well as ESI\dag\ Figs. S1 and S2 for examples of different surface structures). For these surfaces we calculate the OER free energy profile on symmetrically inequivalent titanium, ruthenium, lanthanum and strontium reaction sites in the topmost formula-unit layer - meaning that for example on a defective \ce{BO2} terminated surface also an A atom could become the reaction site.

We consider here only the conventional OER mechanism with four PCET steps (see equation \ref{eq:1}-\ref{eq:4}) that is depicted in Fig. \ref{fig:fig1}b) as we want to study the influence of different reaction sites irrespective of the reaction mechanism. This conventional mechanism does normally not include oxygen ions that are part of the slab. However, in the vicinity of defects and adsorbates it is not always clear if an oxygen originated as part of the adsorbate or the slab and whether the conventional mechanism depicted in Fig. \ref{fig:fig1}b can be applied or not. Therefore, we also include the OER mechanism shown in Fig. \ref{fig:fig1}c which is mathematically equivalent to the conventional mechanism shown in Fig. \ref{fig:fig1}b but operates on an oxygen deficient surface. This so-called lattice oxygen evolution\cite{fabbri2018oxygen} was experimentally observed\cite{wohlfahrt1987oxygen, fierro2007investigation, diaz2013electrochemical} and also explained by basic thermodynamic concepts\cite{binninger2015thermodynamic}. Alternative lattice oxygen evolution mechanisms have been proposed\cite{grimaud2017activating, yoo2018role}, which are however mathematically different from the conventional mechanism (equations \ref{eq:1}-\ref{eq:4}) and were therefore not included in this study.

In total we calculate the OER on 770 symmetrically inequivalent reaction sites (327 \textit{trans} LTON, 135 \textit{cis} LTON, 176 STO, 132 SRO) on 106 LTON, 35 STO and 39 SRO defective (001) surface models. The larger number of reaction sites on the oxynitride stems from anion-induced symmetry breaking. For roughly two thirds of the reaction sites either the *\ce{OOH} or *\ce{OH} intermediate was not stable and we exclude these cases from our analysis.

We calculate the change in free energy ($\Delta G$) of the four reaction steps (1)-(4) using the computational hydrogen electrode (CHE)\cite{norskov2004origin}, where the energy of a proton and an electron equals half the energy of a hydrogen molecule. As the theoretical overpotential is not dependent on the pH or the potential\cite{man2011universality} we perform our calculations at standard conditions (pH=0, $T=298.15$ K) and $U=0$ V. Zero-point energies (ZPE) and entropies ($S$) of the reaction intermediates were included as detailed elsewhere\cite{valdes2008oxidation}. In contrast to other studies, where ZPE was calculated for reaction intermediates in different environments (bridge vs. top site)\cite{valdes2008oxidation}, we always use the ZPE at the top site. Given that for defective surfaces a full ZPE evaluation would be computationally prohibitively expensive and that previously reported changes in ZPE were minor\cite{valdes2008oxidation}, this approach will yield correct trends for defective surfaces.

We estimate the activity of a specific reaction site by the largest step in its OER free energy profile:
\begin{equation}
	\Delta G^{OER}=max[\Delta G_1^0,\Delta G_2^0, \Delta G_3^0, \Delta G_4^0 ]
\end{equation}
$\Delta G_{1-4}^0$ being the change in free energy of the four OER steps (equations \ref{eq:1}-\ref{eq:4}). The calculated overpotential is given by:
\begin{equation}
\eta^{OER}=(\Delta G^{OER})/e-1.23 V
\end{equation}
where 1.23 V is the potential needed to make all $\Delta G$s equal to zero for an ideal catalyst. The adsorption free energies of the intermediate species $\Delta G_{\ce{O}}$, $\Delta G_{\ce{OH}}$ and $\Delta G_{\ce{OOH}}$ were calculated as follows:
\begin{equation}
\Delta G_{ads}=G_{ads+slab}-G_{slab}-nG_{\ce{H2O}}-mG_{\ce{H2}}
\end{equation}
where free energies include changes in ZPE and $S$ while $n$ and $m$ are stoichiometric coefficients that preserve the number of atoms on both sides of the respective reaction.

We further compare the calculated overpotential $\eta^{OER}$ with the overpotential $\eta_D^{OER}$ predicted by the single descriptor $\Delta G_2^0$: It was established that there exists a universal scaling relation between the adsorption energies of the reaction intermediates *\ce{OH} and *\ce{OOH}, their difference being approximatively 3.2 eV for metal and oxide surfaces irrespective of the reaction site\cite{koper2011thermodynamic}. Since the overpotential for oxides and metals is generally determined by either step 2 ($\Delta G_2^0$) or 3 ($\Delta G_3^0 = 3.2 eV - \Delta G_2^0$), the former is often a suitable descriptor of the OER activity that determines the overpotential\cite{man2011universality}:
\begin{equation}
	\eta_D^{OER}=max[\Delta G_2^0,3.2 eV-\Delta G_2^0]/e-1.23 V,
\end{equation}
where $\Delta G_2^0$ is obtained via the difference in adsorption energies.
\begin{equation}
\Delta G_2^0=\Delta G_{\ce{O}}-\Delta G_{\ce{OH}}
\end{equation}
Under the assumption of an optimal $\Delta G_2^0=\Delta G_3^0=\frac{1}{2} 3.2 eV=1.6 eV$, one arrives at a minimum possible overpotential of (1.6 eV)/e-1.23 V=0.37 V. Deviations of $\Delta G_2^0$ from this ideal value lead to larger single-descriptor overpotentials.

\section*{Results and Discussion}

\begin{figure*}
 \centering
 \includegraphics{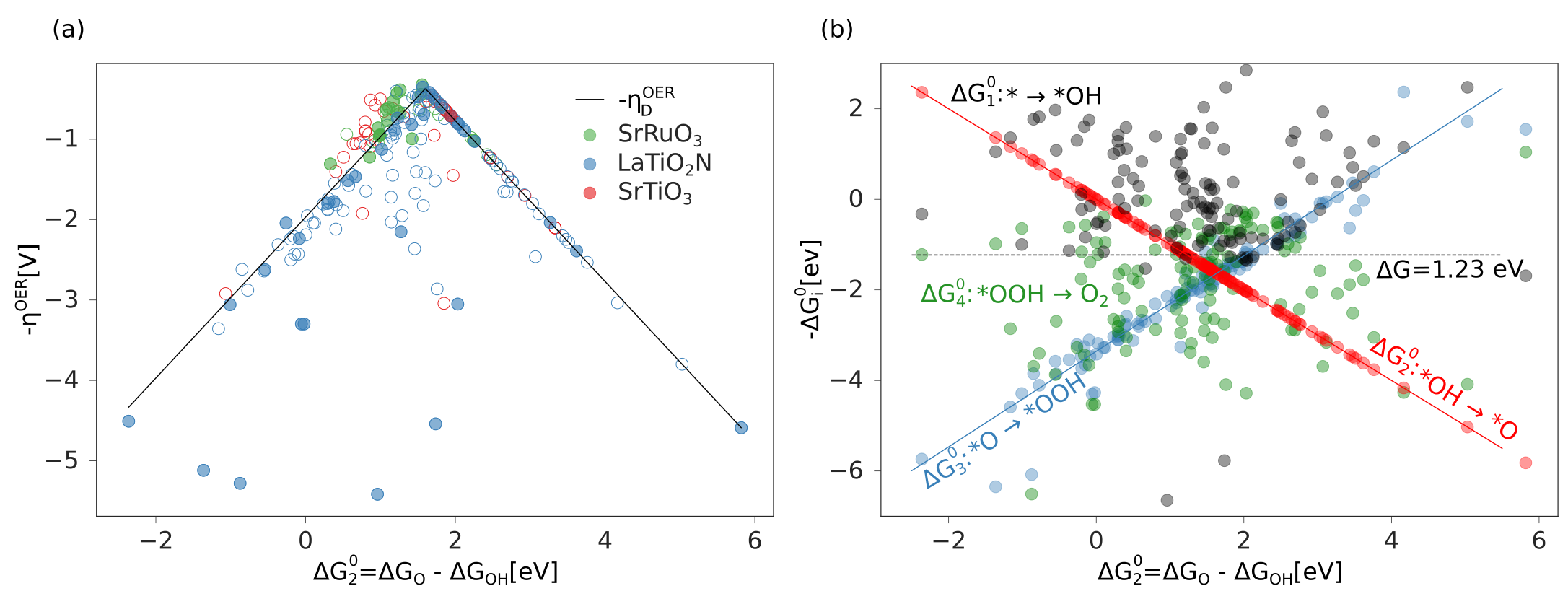}
 \caption{Volcano plot of defective surfaces and analysis of scaling relations of the oxynitride. (a) Computed OER overpotential for defective \ce{LaTiO2N}, \ce{SrRuO3} and \ce{SrTiO3} surfaces plotted against the single descriptor $\Delta G_2^0$, showing the usual volcano shape. Overpotentials for surfaces, which are more likely to occur based on thermodynamic and kinetic arguments (see text) are indicated with filled circles, whereas those that are less likely are indicated with empty circles. (b) Dependence of the negative free energy change of the four OER steps on the descriptor $\Delta G_2^0$ for \ce{LaTiO2N}. The difference between the horizontal $\Delta G = 1.23 eV$ line and the lowest of the four negative reaction energies of a specific OER represents the predicted overpotential. The best linear fit for $-\Delta G_3^0$ is $1.05\Delta G_2^0-3.36$ (Pearson coefficient p=0.91, RSME = 0.64 eV), which is a slight deviation with respect to the scaling law that assumes a slope of 1 and an offset of 3.2 eV.}
 \label{fig:fig2}
\end{figure*}

We begin our analysis by comparing the calculated overpotentials $\eta^{OER}$ with the overpotentials determined via the single descriptor $\eta_D^{OER}$ and find in general a good agreement for the majority of the calculated overpotentials (see Fig. \ref{fig:fig2}a). Most remarkable is the fact that we find for all three materials a continuous distribution of overpotentials within a large range of the descriptor $\Delta G_2^0$, including also the top of the volcano. The complex dissolution kinetics, make it difficult to determine with certainty which sites will be predominantly present on a dissolving surface, yet we can attempt to distinguish between reactive sites occurring more likely (filled circles in Fig. \ref{fig:fig2}a) and less likely (empty circles in Fig. \ref{fig:fig2}a) based on thermodynamic and kinetic considerations: From surface Pourbaix diagrams (see ESI\dag\ section S2 for details) we see, in agreement with experimental studies\cite{adair2006role, connell2012preparation}, that A sites dissolve more readily and therefore B sites represent likelier OER reactive sites. Moreover, the Bell-Evans-Polanyi principle\cite{bell1936theory, evans1936further} implies that the formation of B-site vacancies is also kinetically less likely than the formation of A-site vacancies. Since under OER conditions surfaces are normally oxidized\cite{montoya2014theoretical}, we consider B sites oxidized either by an OH or O species to be the thermodynamically and kinetically most likely OER reaction sites. For LTON we consider the \textit{trans} anion order to be more likely due to electrostatic considerations\cite{ninova2017surface} and for STO we only consider the O covered defect free \ce{TiO2} termination as likely since the material is stable. When considering only the overpotentials of these likely reactive sites we still find a continuous distribution of overpotentials, many of them at the top of the activity volcano. Since many of the less likely (less stable) sites also have overpotentials close to the top of the volcano, this implies, in agreement with recent experiments\cite{roy2018trends}, that there exists no correlation between the stability of a single site and its catalytic activity and that the activity-stability relations must have a different origin.

Given the large number of likely sites close to the top of the volcano, we can assume that some of these will at least be transiently present. Since we omitted combinations of different defects types and considered only zero or full adsorbate coverage, the actual variety of reaction sites should be even larger. Together with alternative reaction mechanisms that could be active, we thus expect even more sites with small overpotentials on a metastable catalyst. Since the activity of a catalyst depends exponentially on the overpotential, even a small number of highly-active sites can dominate the apparent activity\cite{seitz2016highly, sun2018metastable, batchelor2019high}. We therefore postulate that in operando, surface dissolution can lead to the appearance of highly active sites, resulting in an enhanced apparent activity of the catalyst. In other words, the less a material is stable, the higher should be the diversity of reaction sites and the higher the chance of finding highly active reaction sites, which we believe to be the origin of the activity-stability relations. We stress that at this point the proposed activity improvement requires the material to be metastable under the chosen conditions. However, metastable materials such as LTON and SRO can of course only benefit from this proposed activity enhancement if the dissolution is slow enough for the material to still sustain extended periods of operation. 

While previous studies\cite{chang2015activity, danilovic2014activity, chang2014functional} suggest that active dissolution of the material results in the best activities, our results suggest that an activity enhancement can also be achieved by an initial dissolution (preconditioning) of the catalyst that creates the variety of reaction sites but that the dissolution does not have to continue for high OER activities. This agrees with recent experimental findings for \ce{RuO2} thin films, which show that while the surface orientation with the initially highest dissolution rate results in the best activities, an ongoing dissolution is not necessary for highly active surfaces\cite{roy2018trends}.

The similarity of the overpotential distribution of LTON, SRO and STO implies further that the high activity does not originate purely from the bulk electronic structure of the catalyst as is often assumed, but that reaction sites with geometries that enable a lower overpotential can form on any of the investigated materials. This is supported by the experimental finding that perovskite oxide electrocatalysts with structural flexibility can lead to superior activity\cite{fabbri2017dynamic}. We were not able to correlate catalytic activity with geometrical descriptors or simple structural features such as the reaction site species, angles and bond lengths of reaction intermediates or the adsorbate coverages to mention only a few. We also find no correlation with global properties such as the dipole moment of the slab or the nominal charge of the atoms. However, in agreement with other studies\cite{dickens2019electronic, vojvodic2011optimizing, norskov2011density}, a preliminary analysis that will be reported elsewhere\cite{vonruti2019descriptors} hints at a correlation of $\Delta G_2^0$ with the electronic structure and in particular the centre of the O2\textit{p} band of the oxygen adsorbate.

We continue by predicting the minimum overpotentials of the three materials. From the relation between the adsorption energies of the *\ce{OH} and *\ce{OOH} intermediates, we predict minimum overpotentials of 0.23 V for STO, 0.26 V for SRO and 0.40 V for LTON, the two former being smaller than the 0.37 V predicted by the scaling relations, while the latter is larger. Despite these differences, we do for STO and SRO not find actual reaction sites with overpotentials smaller than 0.37 V. This is in line with the observation that breaking the universal scaling relation between *\ce{OOH} and *\ce{OH} is a necessary but not sufficient condition to optimize the OER\cite{govindarajan2018does}. Further, while there are only small differences between the two oxides that show a rather good agreement between the calculated $\eta^{OER}$ and the single-descriptor overpotentials $\eta_D^{OER}$, for LTON we find a not insignificant number of reaction sites that show large deviations from $\eta_D^{OER}$. A more detailed analysis of the reaction free energies of the four charge transfer steps $\Delta G_i^0$ for LTON (see Fig. \ref{fig:fig2}b), reveals that while for most reactions $\Delta G_2^0$ or $\Delta G_3^0$ (blue and red data points) have the largest free-energy change, there are some reactions where $\Delta G_4^0$ (green data points) has the largest free energy change and is the limiting step. This implies that for the oxynitride the step from *OOH to \ce{O2} is energetically less favourable while at the same time the formation of *OH is more favourable. This difference with respect to oxides\cite{man2011universality} suggests a stronger adsorption of *OOH on the oxynitride and leads to larger calculated overpotentials compared to the single-descriptor overpotentials $\eta_D^{OER}$ for some of the OERs.

From the volcano plot in Fig. \ref{fig:fig2}a we can see a further difference between the two oxides and the oxynitride: While for the oxynitride the descriptor value $\Delta G_2^0$ ranges from -2 eV to 6 eV, the two oxides seem to have a much smaller range of $\Delta G_2^0$ and their overpotentials are therefore more concentrated around the top of the volcano. The spread in $\Delta G_2^0$ is important for the apparent activity. A catalyst with an activity of the perfect surface far from the top of the volcano would benefit from a large spread to increase the number of highly active sites. An already good catalyst located not far from the top of the volcano, on the other hand, would benefit from a small spread to have a large proportion of highly active sites. We can explain the difference in the $\Delta G_2^0$ range between oxides and oxynitrides by analysing the charge transfer towards different reaction intermediates. To do so, we look at the change of average L\"owdin charges\cite{lowdin1950non} of the anions for different reaction steps. We find the best correlations of the L\"owdin charges with $\Delta G_2^0$ if we include all anions and not only surface atoms. We compute differences between element-averaged L\"owdin charges for the different reaction steps:
\begin{equation}
	\Delta L_a^{i-j}=\frac{1}{N_a^i}\sum_{x \in a} l_x^i - \frac{1}{N_a^j} \sum_{x \in a} l_x^j
\end{equation}
where $a$ is an anion element (here \ce{O} or \ce{N}), $i$ and $j$ are two consecutive reaction intermediates, $N_a^{i/j}$ is the number of $a$ atoms in the surface structure with the $i/j$ intermediate and $l_x^{i/j}$ is the L\"owdin charge for atom $x$ in the structure with intermediate $i/j$, while the sum runs over all atoms of element $a$.

\begin{figure*}
 \centering
 \includegraphics{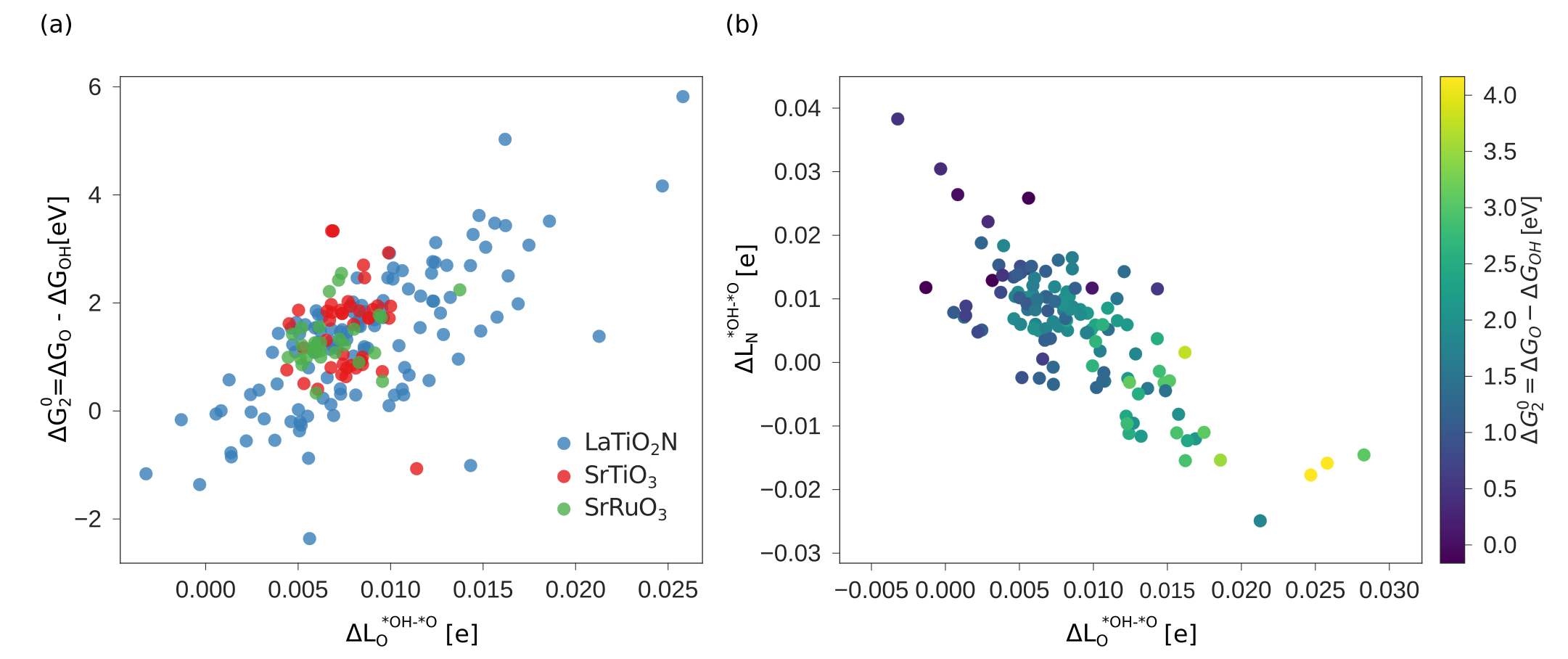}
 \caption{Charge-transfer effects on electrochemistry. (a) Correlation between the change in L\"owdin charge $\Delta L_{\ce{O}}^{*\ce{OH}-*\ce{O}}$ on oxygen and the single descriptor $\Delta G_2^0$. Larger $\Delta L_{\ce{O}}^{*\ce{OH}-*\ce{O}}$ indicate that the oxygen atoms lose more electrons when the reaction intermediate changes from *\ce{OH} to *\ce{O}. This is energetically costly and results in large free energy changes $\Delta G_2^0$ (Pearson coefficient p=0.7 and RSME = 0.89 eV - determined for the oxynitride only). (b) Correlation between the change in L\"owdin charge $\Delta L_{\ce{O}}^{*\ce{OH}-*\ce{O}}$ on oxygen and $\Delta L_{\ce{N}}^{*\ce{OH}-*\ce{O}}$ on nitrogen (p=-0.76). The colour bar indicates the value of the descriptor $\Delta G_2^0$.}
 \label{fig:fig3}
\end{figure*}

As we show in Fig. \ref{fig:fig3}a, there is a direct correlation between the descriptor $\Delta G_2^0$ and the average change in L\"owdin charge of the oxygen ions $\Delta L_O^{*OH-*O}$ for LTON. We do, however, not see a correlation for STO or SRO, which may be related to the smaller range of $\Delta G_2^0$ for these materials. $\Delta L_O^{*OH-*O}$ is generally positive, meaning that the average L\"owdin charge of oxygen is higher for the *\ce{OH} than for the *\ce{O} intermediate, which we can relate to the fact that oxygen in *\ce{OH} attracts electrons from the hydrogen atom and the slab whereas *\ce{O} attracts electrons only from the slab. A smaller $\Delta L_O^{*OH-*O}$ signifies less charge transfer from the slab to the adsorbate during deprotonation, implying that both adsorbates attract a similar number of electrons, which results in an energetically more favourable deprotonation and a smaller $\Delta G_2^0$, giving a physical interpretation to the correlation in Fig. \ref{fig:fig3}a.
 
For the oxides SRO and STO the range of $\Delta L_O^{*OH-*O}$ is much smaller than for LTON leading to a smaller range in $\Delta G_2^0$. We can explain this difference between oxides and oxynitrides by relating the average change in L\"owdin charge of oxygen $\Delta L_O^{*OH-*O}$ to the one of nitrogen $\Delta L_N^{*OH-*O}$. As shown in Fig. \ref{fig:fig3}b we find an inverse correlation between these charge differences, suggesting a charge transfer from N to O. From the colour coding of the data points, we see that $\Delta G_2^0$ is generally smaller when nitrogen loses more electrons (large positive $\Delta L_N^{*OH-*O}$), while the charge variation on the oxygen atoms is small. From these observations we propose that nitrogen can act as an electron reservoir: The reaction from *\ce{OH} to *\ce{O} is favoured if nitrogen ions provide electrons, resulting in a lower descriptor value $\Delta G_2^0$, while reaction sites, where this charge transfer is not possible, result in larger $\Delta G_2^0$s. More generally, this implies that a more flexible valence-band electronic structure, for example in mixed-anion materials, can yield a larger variety of $\Delta G_2^0$s.

\section*{Conclusions}
In summary, we have shown for three electronically different perovskite materials that the structural variety of reaction sites on dissolving heterogeneous catalysts will, for some sites, lead to overpotentials close to the top of the activity volcano. Given that reaction sites with low overpotentials will dominate the apparent activity, this implies that, independent of the bulk electronic properties, operating a heterogeneous catalyst at the border of its stability region can lead to an increased apparent activity due to the increased variety of geometrically different reaction sites. Also preconditioning catalysts under metastable conditions before operation under stable conditions should result in the same activity enhancement.

While we find the single-descriptor approach to generally work well for these highly defective surfaces, we observe that the strong binding of the *\ce{OOH} and *\ce{OH} intermediates on \ce{LaTiO2N} can lead to different limiting steps and larger deviations from the descriptor-based overpotential than for oxides. We find only minor differences in activity between metallic \ce{SrRuO3} and insulating \ce{SrTiO3} but show that a flexible valence band structure as it occurs in \ce{LaTiO2N} yields a larger variety of $\Delta G_2^0$ values.

\section*{Methods}

We determine energies of the various adsorbate covered, defective surfaces by density functional theory (DFT) calculations using the {\sc{Quantum ESPRESSO}}\cite{giannozzi2009quantum} package at the GGA+$U$ level of theory with the PBE\cite{perdew1996generalized} exchange-correlation functional and a Hubbard $U$\cite{anisimov1991band} of 3 eV applied to the titanium 3\textit{d} states. We do not apply a Hubbard $U$ on ruthenium 4\textit{d} orbitals as this setup results in the best agreements with experimentally measured magnetic moments and the density of states at the Fermi energy. We use ultrasoft pseudopotentials\cite{vanderbilt1990soft} with La(5s,5p,5d,6s), Sr(4s,4p,5s), Ti(3s,3p,3d,4s), Ru(4d,5s,5p), O(2s,2p) and N(2s,2p) electrons as valence states to describe the interaction between electrons and nuclei and perform spin-polarized calculations in the case of \ce{SrRuO3}. The cutoff for the plane-wave basis set is 40 Ry and 320 Ry for the kinetic energy and the augmented density respectively for \ce{LaTiO2N} and \ce{SrTiO3}, while for \ce{SrRuO3} we use slightly higher cutoffs of 50 Ry and 500 Ry respectively. We start our calculations from a 40-atom pseudo-cubic perovskite cell and create asymmetric $1\times 1\times 2$ (001) surface slabs (see Fig. \ref{fig:fig1}a) with at least 10 \AA\ vacuum, two fixed atomic layers at the bottom of the slab and a dipole correction in the vacuum layer\cite{bengtsson1999dipole}. The Brillouin zone is sampled with a $4\times 4\times 1$ Monkhorst-Pack\cite{monkhorst1976special} k-point grid for \ce{LaTiO2N} and \ce{SrTiO3} and a $6\times 6\times 1$ grid for \ce{SrRuO3}. We relax ionic positions until forces converge below 0.05 eV\AA$^{-1}$ and total energies change by less than $1.4\cdot10^{-5}$ eV.

\section*{Conflicts of interest}
There are no conflicts to declare.

\section*{Acknowledgements}
This research was funded by the SNF Professorship Grant PP00P2\_157615. Calculations were performed on UBELIX (http://www.id.unibe.ch/hpc), the HPC cluster at the University of Bern, the Swiss National Supercomputing Centre (CSCS) under project s766 and SuperMUC at GCS@LRZ, Germany, for which we acknowledge PRACE for awarding us access.

\balance

\bibliography{library}


\clearpage
\clearpage 
\setcounter{page}{1}
\renewcommand{\thetable}{S\arabic{table}}  
\setcounter{table}{0}
\renewcommand{\thefigure}{S\arabic{figure}}
\setcounter{figure}{0}
\renewcommand{\thesection}{S\arabic{section}}
\setcounter{section}{0}
\renewcommand{\theequation}{S\arabic{equation}}
\setcounter{equation}{0}
\onecolumngrid

\begin{center}
\textbf{Supplementary information for\\\vspace{0.5 cm}
\large The role of metastability in enhancing water-oxidation activity\\\vspace{0.3 cm}}
Nathalie Vonrüti and Ulrich Aschauer

\small
\textit{Department of Chemistry and Biochemistry, University of Bern, Freiestrasse 3, CH-3012 Bern, Switzerland}

(Dated: \today)
\end{center}

\section{Surface examples}

\begin{figure*}[h]
 \centering
 \includegraphics[width=0.4\textwidth]{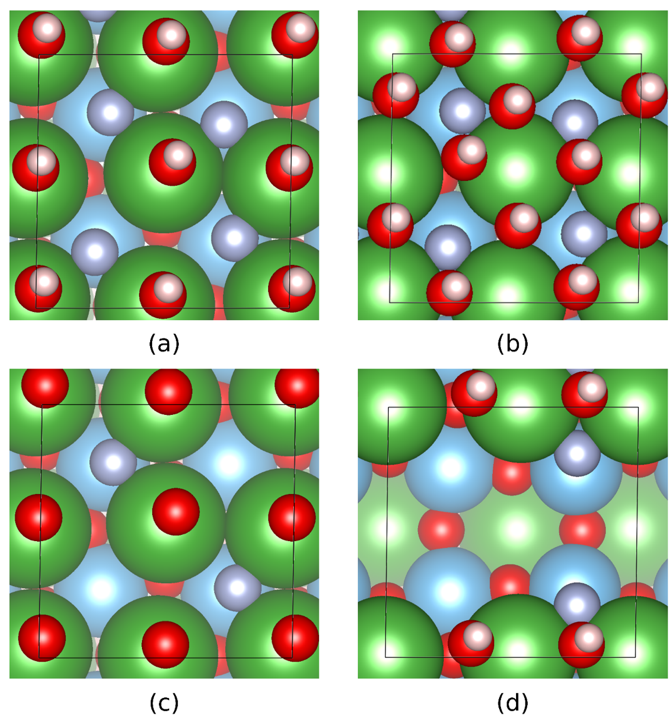}
 \caption{Top view of different A terminated defective surfaces (images show atoms before relaxation) for the example of \textit{trans} LaTiO$_2$N. (a) No defects, A site covered with OH in top position (b) No defects, OH adsorbates in bridge position between A sites (c) two diagonal nitrogen vacancies, O adsorbates in the top position (d) two linear nitrogen vacancies and two linear A vacancies, OH adsorbates between remaining A sites in bridge position. Colour code: La (A-site) = green, Ti (B-site) = blue, O = red, H = white.}
 \label{fig:SI1}
\end{figure*}

\begin{figure*}[h]
 \centering
 \includegraphics[width=0.4\textwidth]{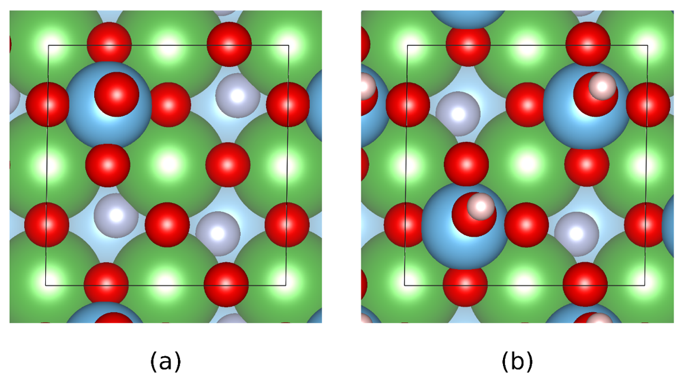}
 \caption{Top view of different B terminated defective surfaces (images show atoms before relaxation) for the example of \textit{trans} LaTiO$_2$N. (a) three B vacancies, remaining B site covered with O in top position (b) two diagonal B vacancies, remaining B sites covered with OH in top position. Colour code: La (A-site) = green, Ti (B-site) = blue, O = red, H = white.}
 \label{fig:SI2}
\end{figure*}

Figures \ref{fig:SI1} and \ref{fig:SI2} show examples of A terminated and B terminated surfaces respectively, depicting some representative adsorbate and defect configurations.
 
\section{Occurrence of surface structures - partial surface Pourbaix diagrams}

To qualitatively assess the likelihood for different defective surfaces to occur, we calculate partial surface Pourbaix diagrams – meaning that we analyse the potential and pH dependent occurrence of different vacancy types independently of each other. A conventional surface Pourbaix diagram would only show that our metastable materials are not thermodynamically stable for relevant potentials and pHs. However, by comparing the stability of surfaces with single defects to the perfect stoichiometric surface we obtain qualitative information as to which surface structures are more likely to be present than others.

For these calculations we consider defect formation accompanied by formation of solvated ionic species. Since the energy of these solvated ionic species is difficult to calculate within DFT, we use the approach of Persson et al. \citeSI{persson2012predictionSI}, where the chemical potentials are calculated with respect to experimental reference energies, allowing to combine DFT energies and Gibbs free energies of solvated ions. We also apply a correction to the energy of the oxygen atom to correct for the well-known overbinding of O$_2$ within DFT. We obtain the correction of 0.68 eV via fitting the experimental formation enthalpies of metal oxides (no transition-metal oxides) to our DFT energies\citeSI{wang2006oxidationSI}. Further, DFT+$U$ calculations impose challenges in calculating formation enthalpies of reactions involving compounds with both localized and delocalized electronic states. Therefore, we also add a correction to every element to which a Hubbard $U$ was applied\citeSI{jain2011formationSI} such as to correctly reproduce the formation enthalpy of TiO$_2$ (for Ru no correction is applied as we use no Hubbard $U$ in our calculations). As we do not include surface adsorbates in this analysis, we do not include zero-point energies of adsorbates. Finally, to be able to determine the partial surface Pourbaix diagram with a numerical solver, we prevent the dissolution of the surface by imposing the presence of atoms that are part of the so-called reference set.

\begin{figure*}[h]
 \centering
 \includegraphics{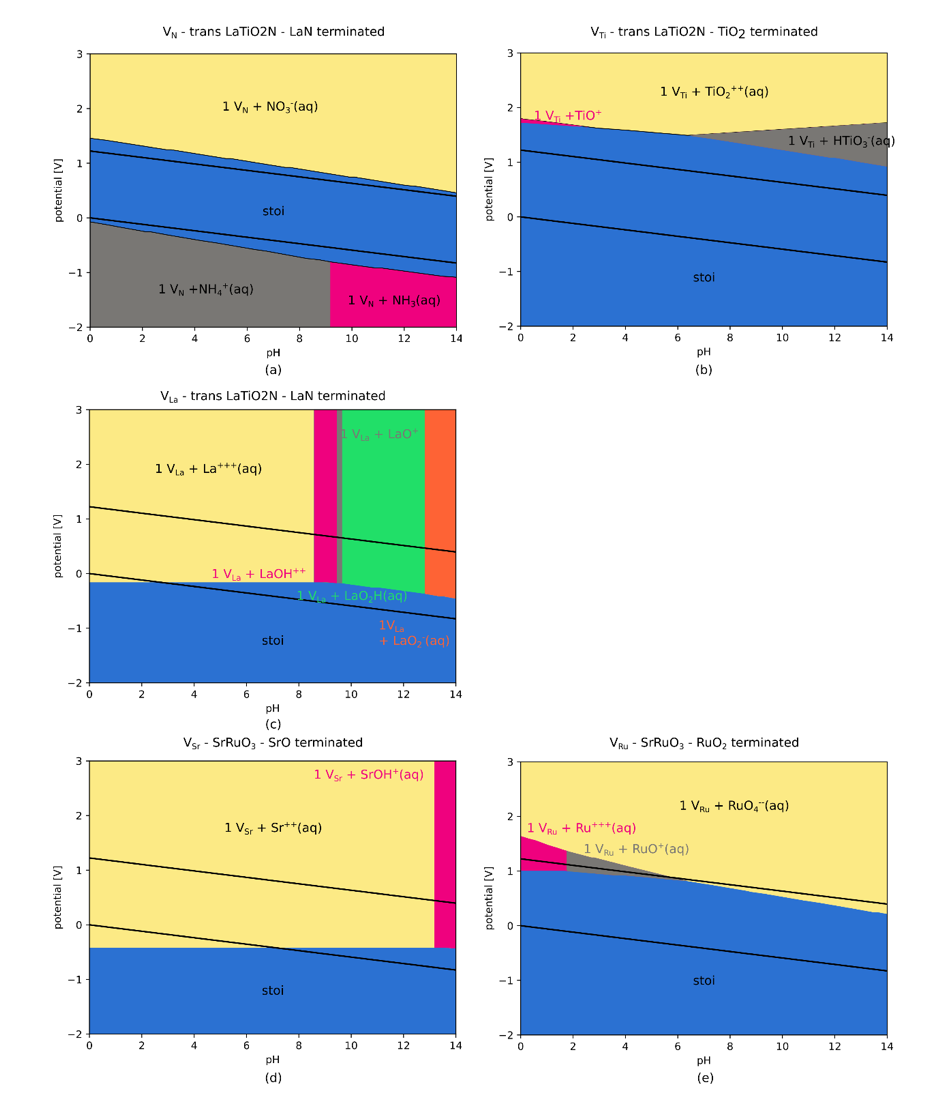}
 \caption{Partial surface Pourbaix diagrams, each containing the perfect stoichiometric surface termination and a surface structure with one specific vacancy only. The dark blue phase in the Pourbaix diagram represents the area where the perfect stoichiometric surface is stable. The black lines surround the stability region of water. (a) LaN-terminated \textit{trans} LaTiO$_2$N surface with one V$_\mathrm{N}$, (b) TiO$_2$-terminated \textit{trans} LaTiO$_2$N surface with one V$_\mathrm{Ti}$, (c) LaN-terminated \textit{trans} LaTiO$_2$N surface with one V$_\mathrm{La}$, (d) SrO-terminated SrRuO$_3$ surface with one V$_\mathrm{Sr}$ and (e) RuO$_2$-terminated SrRuO$_3$ surface with one V$_\mathrm{Ru}$.}
 \label{fig:SI3}
\end{figure*}

Figure \ref{fig:SI3} shows the obtained partial surface Pourbaix diagrams where we consider only the stoichiometric perfect surface (stoi) and a single vacancy. We do not consider oxygen vacancies as these are likely to be spontaneously healed by O* adsorbates under OER conditions\citeSI{ouhbi2019nitrogenSI}. From Figures \ref{fig:SI3} a)-c), we see that at a given pH La vacancies (depending on the pH via formation of various solvated species) on the LaTiO$_2$N surface start appearing already at potentials below the equilibrium potential for water oxidation. Nitrogen vacancies start forming slightly above the water oxidation potential, whereas Ti vacancies require the highest potential to form. A similar trend is observed for SrRuO$_3$ in Figures \ref{fig:SI3} d)-e), where Sr vacancies are formed at significantly lower potentials than Ru vacancies. From this analysis it thus seems that for our set of materials A-site vacancy formation occurs at smaller potentials than B-site vacancy formation, while the potential for N vacancy formation is similar to the B-site. This agrees with experimental observations that the B-site in perovskite oxides is generally thermodynamically more stable than the A-site in contact with an aqueous solvent\citeSI{adair2006roleSI, connell2012preparationSI}.

The thermodynamic stability is generally correlated with the kinetic stability via the Bell-Evans-Polanyi principle\citeSI{bell1936theorySI, evans1936furtherSI}, which states that kinetic activation barriers ($E_a$) obey the relation $E_a=E_0+\alpha\Delta H$, where $E_0$ is a reference energy for a given class of reactions, $\Delta H$ is the thermodynamic reaction enthalpy and $\alpha$ characterises the position of the transition state along the reaction coordinate. This principle implies that reactions with a larger thermodynamic energy difference ($\Delta H$) will also have a larger activation barrier. Applied to our dissolving surfaces, this implies that the formation of B-site vacancies is both thermodynamically and kinetically less likely than the formation of A-site vacancies. We thus consider that surfaces with B reaction sites are more likely to exist on both thermodynamic and kinetic grounds than surfaces with A reaction sites.

\clearpage
\bibliographystyleSI{apsrev4-1}
\bibliographySI{library}

\end{document}